\documentclass[conference, a4paper]{IEEEtran}
\IEEEoverridecommandlockouts

\usepackage{cite}
\usepackage{amsmath,amssymb,amsfonts}
\usepackage[ruled,vlined]{algorithm2e}

\usepackage{algpseudocode}
\usepackage{hyperref}
\usepackage{cleveref}

\usepackage{array}
\usepackage{graphicx}
\usepackage{textcomp}
\usepackage{xcolor}
\usepackage{hyperref}
\usepackage{tabularray}
\def\BibTeX{{\rm B\kern-.05em{\sc i\kern-.025em b}\kern-.08em
    T\kern-.1667em\lower.7ex\hbox{E}\kern-.125emX}}

\begin{document}

\title{An Interference-aware Approach for Co-located Container Orchestration with Novel Metric\\
\thanks{This work is supported by National Key R\&D Program of China (No.2021YFB3300200), the National Natural Science Foundation of China (No. 62072451, 62102408), Shenzhen Industrial Application Projects of undertaking the National key R \& D Program of China (No. CJGJZD20210408091600002), and Shenzhen Science and Technology Program (No. RCBS20210609104609044). Minxian Xu is the corresponding author.}
}

\author{\IEEEauthorblockN{Xiang Li$^{1, 2}$, Linfeng Wen$^{1, 2}$, Minxian Xu$^{1}$, Kejiang  Ye$^{1}$}
\IEEEauthorblockA{1. Shenzhen Institute of Advanced Technology, 
Chinese Academy of Sciences, 
Shenzhen, China\\
2. University of Chinese Academy of Sciences, China\\
\{xiang.li1, lf.wen, mx.xu, kj.ye\}@siat.ac.cn}
}

\maketitle

\begin{abstract}


Container orchestration technologies are widely employed in cloud computing, facilitating the co-location of online and offline services on the same infrastructure. Online services demand rapid responsiveness and high availability, whereas offline services require extensive computational resources. However, this mixed deployment can lead to resource contention, adversely affecting the performance of online services, yet the metrics used by existing methods cannot accurately reflect the extent of interference.

In this paper, we introduce scheduling latency as a novel metric for quantifying interference and compare it with existing metrics. Empirical evidence demonstrates that scheduling latency more accurately reflects the performance degradation of online services. We also utilize various machine learning techniques to predict potential interference on specific hosts for online services, providing reference information for subsequent scheduling decisions. Simultaneously, we propose a method for quantifying node interference based on scheduling latency. To enhance resource utilization, we train a model for online services that predicts CPU and MEM (memory) resource allocation based on workload type and QPS. Finally, we present a scheduling algorithm based on predictive modeling, aiming to reduce interference in online services while balancing node resource utilization.
Through experiments and comparisons with three other baseline methods, we demonstrate the effectiveness of our approach. Compared with three baselines, our approach can reduce the average response time, 90th percentile response time, and 99th percentile response time of online services by 29.4\%, 31.4\%, and 14.5\%, respectively.


\end{abstract}

\begin{IEEEkeywords}
container, interference detection, container orchestration, scheduler
\end{IEEEkeywords}

\section{Introduction}
Linux container is a form of operating system-level virtualization used to run multiple isolated user space environments on a single Linux host\cite{chae2019performance}. They enable packaging applications and all their dependencies into a self-contained runtime environment, thus achieving more efficient resource utilization, faster deployment speeds, and enhanced portability. Among the most prominent container technologies is Docker\cite{docker}, but other implementations such as Podman, LXC, and rkt also exist\cite{podman,LXC,rkt}.

Container orchestration refers to the process of managing, orchestrating, and automating the deployment of large-scale containerized applications across one or more hosts\cite{Casalicchio2019}. Online and offline workloads container orchestration involves managing and deploying online (real-time) and offline (batch processing) workloads that serve different purposes within the same container orchestration platform. Online workloads encompasses applications with real-time requirements, such as web applications, real-time analytics, and real-time communication, necessitating rapid responsiveness and high availability. Offline workloads typically comprises batch processing jobs like data processing, report generation, and large-scale data analysis, which do not demand real-time responses but may require substantial computing resources\cite{8258257}. This co-location strategy enhances resource utilization, simplifies management, reduces costs, and ensures both high availability and flexibility.

While co-deploying online and offline workloads on the same container orchestration platform can share resources\cite{doi:10.1177/10943420231167811}, it may also introduce challenges and issues. Importantly, in co-location scenarios, online and offline workloads may compete for computational resources, especially in resource-constrained environments. Offline jobs may consume a significant portion of computational resources, thereby impacting the performance and response times of online workloads. In other words, online workloads are susceptible to interference from offline workloads during co-location. One current challenge is the absence of a general method for quantifying the degree of such interference. Therefore, to guarantee the performance of online workloads, we can minimize interference among containers through container orchestration. For the realization of this approach, detecting interference becomes critically important\cite{9213058}.

In existing methods\cite{6258001}, resource utilization metrics such as CPU usage, memory usage, disk I/O, and network I/O are typically collected. However, due to granularity limitations, these data provide limited information and do not effectively reflect the interference experienced by online workloads. In addition to these metrics, we incorporate hardware-level data, such as branch prediction failures and cache misses. Excessive branch prediction failures or cache misses can introduce time overhead to some extent. Furthermore, we place special emphasis on the metric of "scheduling latency", which refers to the interval between a process entering the scheduling queue after exhausting its CPU time slice and regaining CPU control. We believe this metric can also partially reflect changes in online workloads response times and offer valuable insights.

To address these challenges, this paper introduces a metric that effectively reflects interference and proposes a quantification method for interference. We also train a prediction model to forecast this interference value. Finally, we design a scheduling algorithm and implement a corresponding scheduler to minimize interference during co-location. Through experiments, we validate the effectiveness of our approach. The contributions of this paper can be summarized as follows:

\begin{enumerate}
  \item We introduce a novel metric, scheduling latency, that effectively captures interference.
  \item We propose a quantification method for interference that accurately reflects the interference experienced by online workloads.
  \item We trained a random forest model for predicting the interference online workloads will encounter when deployed on a host.
  \item Building upon the interference quantification method and prediction model, we design a scheduling algorithm and implement a corresponding scheduler to minimize the interference experienced by online workloads during deployment.
\end{enumerate}

The rest of the paper is organized as follows: Section \ref{Motivation} presents the motivation to use the novel metric to evaluate the performance, Section \ref{Related Work} discusses the related work contributing to the relevant areas, Section \ref{System Framework} presents the system framework that the applications can be co-located and analyzed with interference, Section \ref{Experimental Results} demonstrates the experimental results, and finally the Section \ref{Conclusions and Future Work} concludes the whole paper and highlights some future research directions.

\section{Motivation} \label{Motivation}
In the aforementioned sections, we have mentioned scheduling latency, which refers to the time interval between a process entering the scheduling queue due to the exhaustion of its CPU time slice and regaining CPU control. We hypothesize that a larger scheduling latency for online workloads will result in increased response time. This implies that even though the average CPU utilization may be similar, varying scheduling latency may lead to different performance in terms of response time.

To verify whether there is a correlation between the response time of online workloads and both CPU utilization and scheduling latency or not, and if such a correlation exists, which factor is more significant, we conducted two sets of experiments using Web Search\footnote{https://github.com/parsa-epfl/cloudsuite/blob/main/docs/benchmarks/web-search.md} and In-Memory Analytics\footnote{https://github.com/parsa-epfl/cloudsuite/blob/main/docs/benchmarks/in-memory-analytics.md}.

Details of our experimental environment are as follows: all experiments were conducted on machines with 32 cores and 64GB of RAM, running the Ubuntu 22.04.1 LTS operating system. The selected benchmarks included Web Search as an online workload and In-Memory Analytics as an offline workload.

The first set of experiments involved keeping the QPS of Web Search constant (300 requests per second) while varying the CPU utilization of In-Memory Analytics. Initially, we allocated 2 CPU cores for In-Memory Analytics, and after each set of experiments, we incrementally increased the available CPU cores by two, totaling 10 sets of experiments.

And the second set of experiments involved maintaining the number of CPU cores available for In-Memory Analytics constant (8 cores) and modifying the request frequency of Web Search. Initially, the request frequency for Web Search was set at 200 requests per second. After each set of experiments, we increased the request frequency by 200, resulting in 10 sets of experiments.

\begin{figure}[htbp]
  \centering
  \begin{minipage}{0.47\linewidth} 
    \centering
    \includegraphics[width=\linewidth]{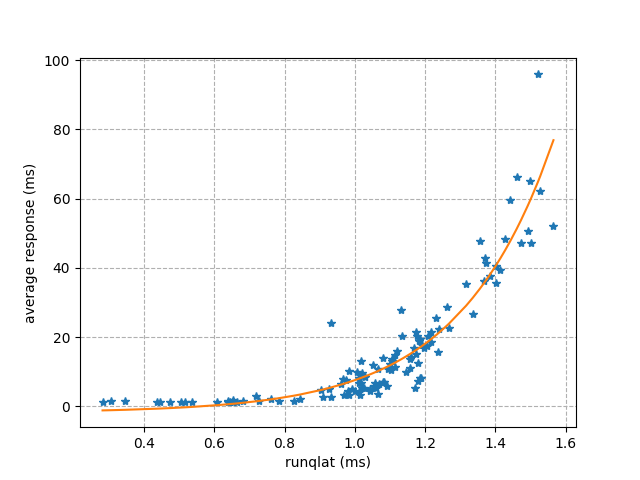}
    \caption{Experiment 1: correlation between average response time and scheduling latency}
    \label{fig:image1}
  \end{minipage}%
    \begin{minipage}{0.06\linewidth}
    \centering
    \qquad
  \end{minipage}%
    \begin{minipage}{0.47\linewidth} 
    \centering
    \includegraphics[width=\linewidth]{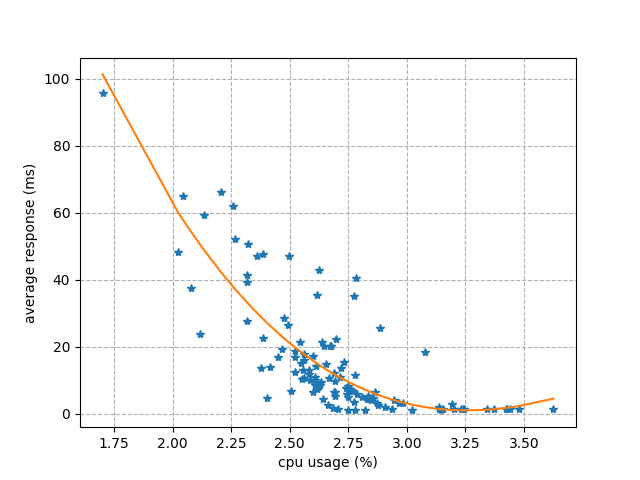}
    \caption{Experiment 1: correlation between average response time and CPU Utilization}
    \label{fig:image2}
  \end{minipage}
\end{figure}

\begin{figure}[htbp]
  \centering
  \begin{minipage}{0.47\linewidth} 
    \centering
    \includegraphics[width=\linewidth]{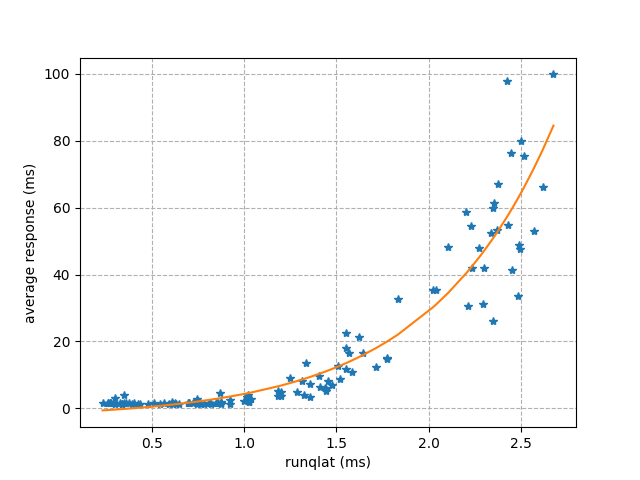}
    \caption{Experiment 2: correlation between average response time and scheduling latency}
    \label{fig:image3}
  \end{minipage}%
  \begin{minipage}{0.06\linewidth}
    \centering
    \qquad
  \end{minipage}%
  \begin{minipage}{0.47\linewidth} 
    \centering
    \includegraphics[width=\linewidth]{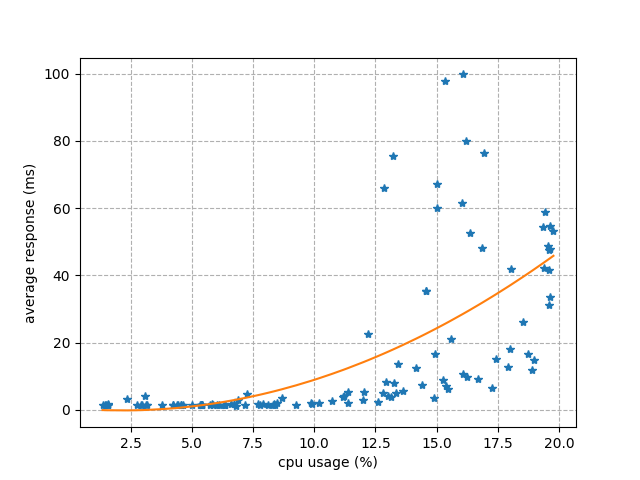}
    \caption{Experiment 2: correlation between average response time and CPU Utilization}
    \label{fig:image4}
  \end{minipage}
\end{figure}

In both sets of experiments, we recorded the relationships between CPU utilization and average response latency, as well as between average scheduling latency and average response latency. As depicted in Figure \ref{fig:image1} and Figure \ref{fig:image2}, the horizontal axis represents the CPU utilization of online workloads, while the vertical axis represents the response time. In Figure \ref{fig:image3} and Figure \ref{fig:image4}, the horizontal axis represents the average scheduling latency within a minute for online workloads, and the vertical axis represents the response time.

For each graph, we attempted to fit a curve to these data points and assessed the quality of the fitness. Data presented in Table \ref{tab:mytable} indicate that the correlation between average scheduling latency and response time is noticeably superior to that between CPU utilization and response time. This suggests that scheduling latency has more impacts on response time.
In general, an increase in the number of processes in the scheduling queue results in a corresponding increase in scheduling latency. If the scheduling latency for online workloads remains consistently high, it typically leads to longer response times.


\begin{table}
\centering
\caption{Curve Fitting Performance}
\label{tab:mytable}
\begin{tblr}{
  cells = {c},
  hline{1,6} = {-}{0.08em},
  hline{2} = {-}{0.05em},
}
Experiments        & MAPE & R²   \\
Exp1. runqlat-resp & 0.46 & 0.88 \\
Exp1. cpu-resp     & 0.77 & 0.68 \\
Exp2. runqlat-resp & 0.60 & 0.88 \\
Exp2. cpu-resp     & 1.31 & 0.42 
\end{tblr}
\end{table}

\section{Related Work} \label{Related Work}

Currently, experts and scholars both domestically and internationally are conducting extensive research in the fields of interference detection and container orchestration within cloud computing environments. These studies aim to address various challenges in cloud computing, including performance optimization, resource allocation, scalability, and security. They actively explore new methods and tools, continuously improving interference detection efficiency, and utilizing container orchestration technology to meet various scope requirements.

\subsection{Interference Detection}

David et al. proposed a system named Heracles\cite{7284086}. They began by analyzing that different types of online workloads exhibit varying degrees of sensitivity to different types of interference. And then they designed and implemented a heuristic feedback-based system resource controller, ensuring that latency-critical online workloads can coexist with batch processing tasks on the same host without violating SLOs. However, this approach is only applicable when deploying a single online workload alongside multiple offline workloads on a host. 


Qiu et al. introduced FIRM\cite{258914}, an intelligent fine-grained resource management framework. It addresses challenges in sharing computational resources across microservices, as contention can lead to delays and breach user request SLOs. FIRM aims to improve overall resource utilization through predictable resource sharing among microservices. However, integrating a reinforcement learning model into FIRM presents challenges, including high training costs and model convergence difficulties.

Chen et al. observed changes in hardware-level metrics when several typical combinations of online and offline workloads were co-located and subsequently recommended certain co-location combinations while discouraging others\cite{8855680}. However, Chen did not propose a quantification method for interference, limiting the generalizability of their approach.

Xu et al. investigated the trade-offs between the dominant scaling techniques, including horizontal scaling, vertical scaling, and brownout in terms of execution cost and response time\cite{9904920}. Their goal is to ensure that microservices systems maintain Quality of Service (QoS) under various workloads through efficient scaling methods. They propose a prediction algorithm based on gradient recurrent units, which helps to accurately predict workload and facilitate efficient scaling, and a multi-faced scaling method using reinforcement learning, which makes scaling decisions.

Luo et al. proposed an efficient resource management system, namely Erms\cite{10.1145/3567955.3567964}. They established response models for each task and calculated the available resources for each container during deployment based on the required SLA. However, Luo's modeling only accounted for potential interference from CPU and memory utilization, and creating separate models for each task incurred excessive computational overhead. 

Xu et al. presented the architecture of Alibaba’s microservice cluster designed to handle large-scale microservice management, along with comprehensive statistical analysis of the microservices in its production environmentl\cite{xu2023practice}. They propose enhanced resource allocation methods that build upon Alibaba’s current practices to efficiently and elastically support services through various means such as workload estimation, capability modeling, and resource allocation policies.

\subsection{Container Orchestration}

Rodriguez et al. conducted research on container orchestration management systems and proposed a classification framework to identify various mechanisms that can address the challenges in this domain\cite{https://doi.org/10.1002/spe.2660}. They then applied this proposed classification to various state-of-the-art systems to identify research gaps and open challenges in the literature, serving as future directions for researchers. Their work primarily focuses on system modeling and design rather than delving into the details of orchestration strategies.

Casalicchio et al. surveyed advanced container technologies\cite{https://doi.org/10.1002/cpe.5668}. They introduced fundamental container concepts like images and Docker, discussed how to containerize applications, and explored container orchestration with tools like Kubernetes and Docker Swarm. The paper also covered container security, addressing challenges in isolation, access control, image security, and vulnerability management. It concluded by suggesting future research directions.


Rodriguez et al. proposed a comprehensive container resource management approach\cite{rodriguez2020container} with three key objectives: optimizing initial container placement, dynamically adjusting resource allocation based on cluster workload, and improving resource efficiency through rescheduling when possible. However, the heuristic resource scheduling method employed in this study may struggle to adapt to dynamic environments, resulting in a decrease in performance.

Struhár et al. proposed a container orchestration method for real-time systems\cite{9613685}, aiming to prevent excessive CPU resource allocation during container scheduling. They also addressed weak isolation issues in container-based virtualization. They proposed performance metrics for both container and node levels, useful for admission control and real-time behavior adjustments in container deployment. They implemented these ideas on Kubernetes, but their approach primarily focuses on real-time systems without considering future changes.

Compared to existing work, our paper conducts a more comprehensive study on service interference. Specifically, we no longer rely solely on heuristic methods; instead, we train a model to predict interference values. Additionally, we introduce an effective interference metric and propose a quantification method for interference in an innovative manner. Finally, we have designed and implemented a novel scheduling algorithm that better aligns with our objectives.

\section{System Framework} \label{System Framework}

\begin{figure}[htbp]
  \centering
  \includegraphics[width=0.9\linewidth]{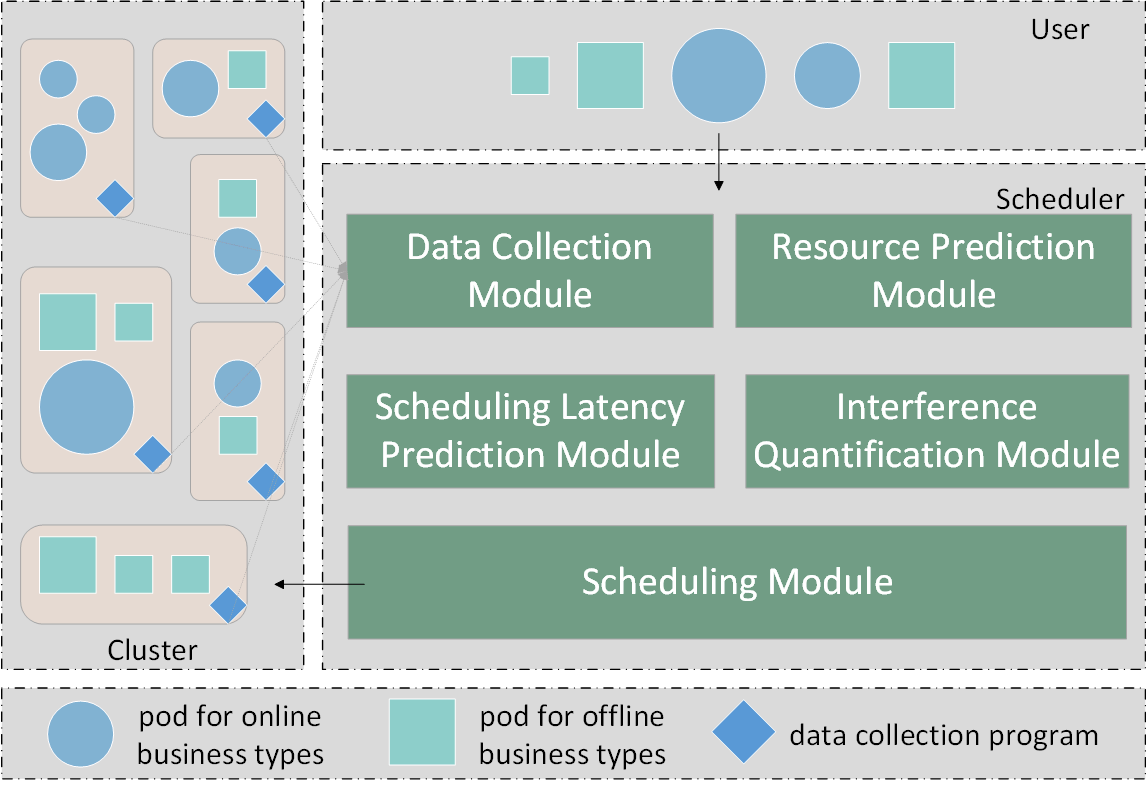}
  \caption{System Framework}
  \label{fig:framework}
\end{figure}

In this section, we will introduce the system framework that we have implemented. As depicted in Figure \ref{fig:framework}, the workflow of this system is as follows: For each pod submitted by a user, it first undergoes resource allocation, including CPU and memory resources, determined by the \textit{Resource Prediction Module}. Subsequently, within the \textit{Interference Quantification Module}, two primary tasks are accomplished. The first task involves predicting the scheduling latency for deploying the pod on each node, based on performance data collected by the \textit{Data Collection Module} and the user-specified QPS (Queries Per Second), using the \textit{Scheduling Latency Prediction Module}. The interference level of the pod is calculated based on the prediction results. The second task is to compute the interference level for each node. Finally, in the \textit{Scheduling Module}, the node allocation for the pods is determined based on the results from the \textit{Interference Quantification Module}, completing the deployment process.

The system framework comprises five primary modules, each responsible for a specific task, which will be elaborated on in the following sub sections.



\subsection{Data Collection Module}
On each node of the cluster, a data collection program is running, primarily tasked with real-time collection of application-layer data, hardware-layer data, and scheduling latency data from the host. Additionally, this program listens on a port, and upon receiving a request from the Data Tracer through this port, promptly provides the current host's data in response.

\subsection{Resource Prediction Module}
The responsibility of the \textit{Resource Prediction Module} is to forecast resources for submitted pods. Specifically, based on the provided request frequency and task type, it is primarily tasked with predicting CPU usage and memory usage.

\begin{figure}[htbp]
    \centering
  \begin{minipage}{0.47\linewidth} 
    \centering
    \includegraphics[width=\linewidth]{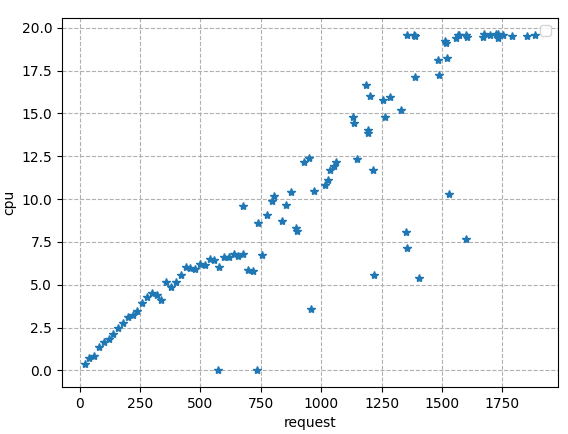}
    \caption{Correlation between QPS and CPU Utilization}
    \label{fig:request-cpu}
  \end{minipage}%
    \begin{minipage}{0.06\linewidth}
    \centering
    \qquad
  \end{minipage}%
  \begin{minipage}{0.47\linewidth} 
    \centering
    \includegraphics[width=\linewidth]{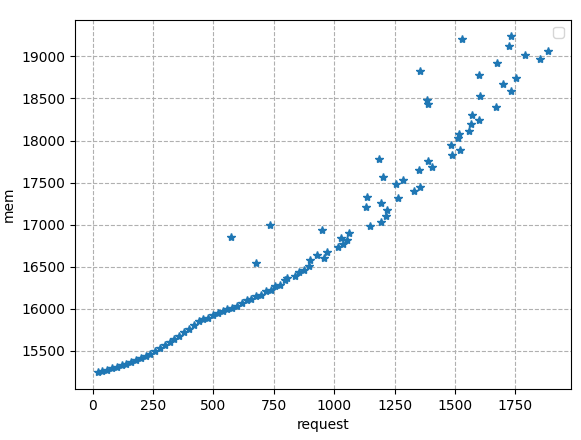}
    \caption{Correlation between QPS and Memory Usage}
    \label{fig:request-mem}
  \end{minipage}
\end{figure}

We conducted a statistical analysis of CPU utilization and memory consumption under different request frequencies, and the relevant data is presented in Figure \ref{fig:request-cpu} and Figure \ref{fig:request-mem}. These data exhibit a clear linear relationship. Therefore, we employed a linear regression model to predict CPU utilization and memory usage for applications.

\subsection{Scheduling Latency Prediction Module}
\label{subsec:runqlat-module}
Our aim is to develop a model that can predict online workloads scheduling latency based on the following factors: the type of online workloads, request volume, host resource utilization (such as CPU utilization, memory utilization, network I/O, disk I/O, etc.), host hardware data, and the distribution of scheduling latency for various processes within the host. Simultaneously, our goal is to ensure that the model is light-weight enough to be encapsulated as a single component for use by other applications, such as integration into schedulers.

\begin{figure}[h]
  \centering
  \includegraphics[width=1\linewidth]{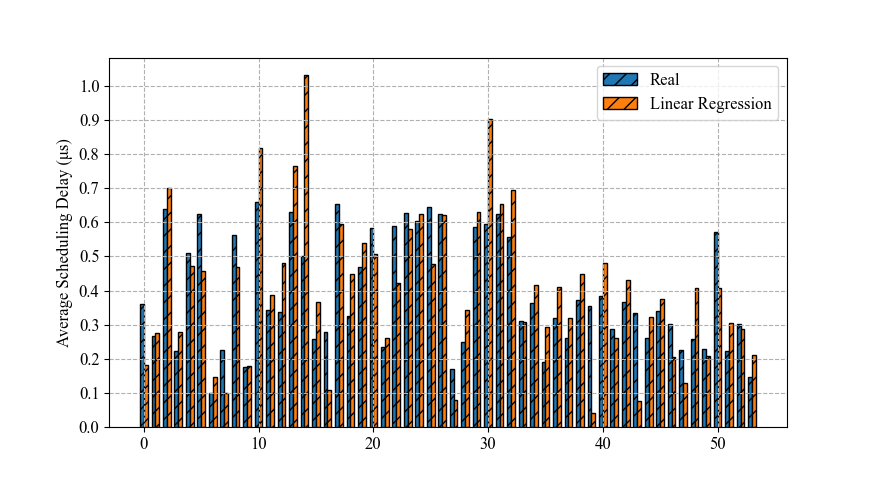}
  \caption{Comparison of Linear Regression Model Predictions}
  \label{fig:image5}
\end{figure}

\begin{figure}[h]
  \centering
  \includegraphics[width=1\linewidth]{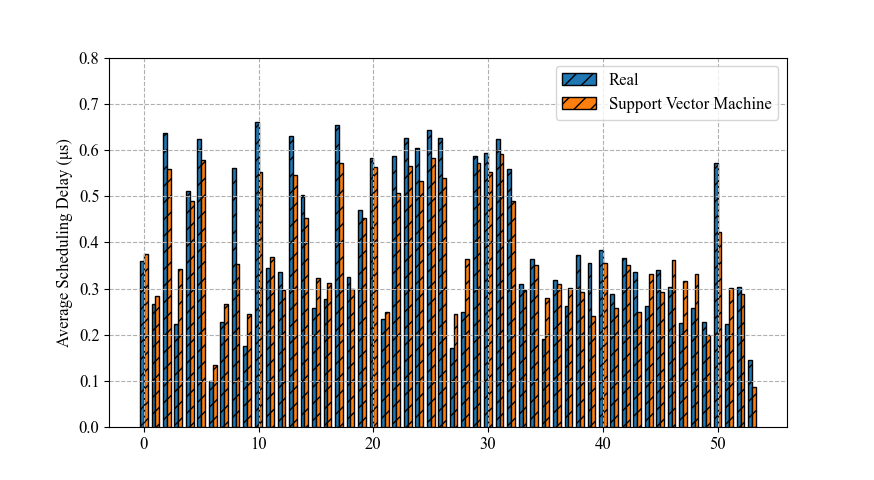}
  \caption{Comparison of Support Vector Machine Model Predictions}
  \label{fig:image6}
\end{figure}

\begin{figure}[h]
  \centering
  \includegraphics[width=0.95\linewidth]{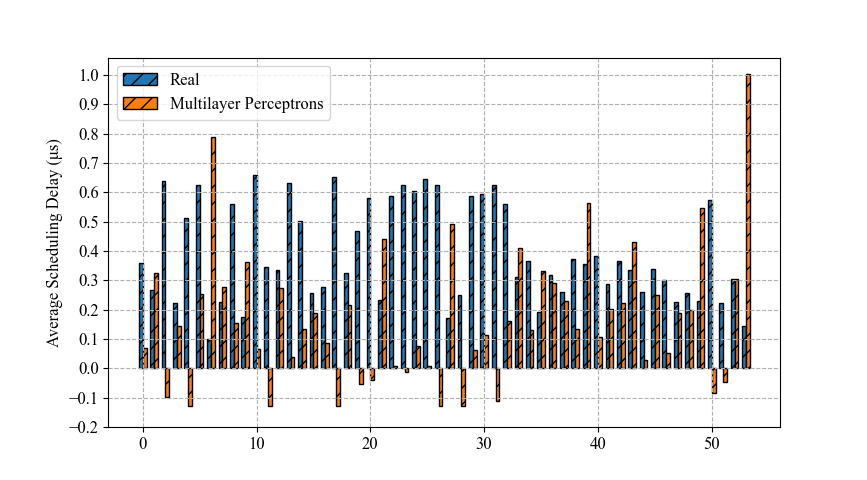}
  \caption{Comparison of Multilayer Perceptron Model Predictions}
  \label{fig:image7}
\end{figure}

\begin{figure}[h]
  \centering
  \includegraphics[width=0.95\linewidth]{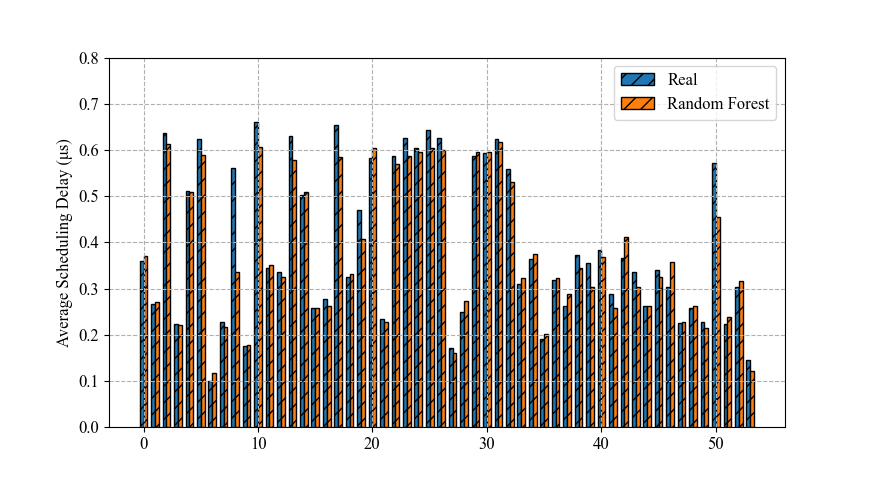}
  \caption{Comparison of Random Forest Model Predictions}
  \label{fig:image9}
\end{figure}

\begin{figure}[h]
  \centering
  \includegraphics[width=1\linewidth]{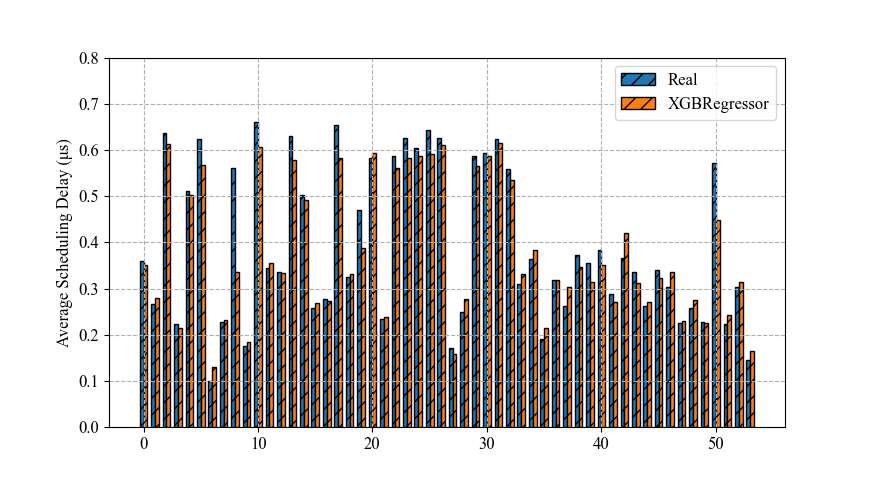}
  \caption{Comparison of XGBRegressor Model Predictions}
  \label{fig:xgboost}
\end{figure}

To achieve these objectives, we conducted a series of experiments employing common machine learning methods, including Linear Regression\cite{su2012linear}, Support Vector Machine\cite{suthaharan2016machine}, Multilayer Perceptron\cite{ramchoun2016multilayer},  Random Forest\cite{biau2016random}, and XGBRegressor\cite{chen2016xgboost}. The experimental results are shown in Figures \ref{fig:image5}, \ref{fig:image6}, \ref{fig:image7}, \ref{fig:image9}, \ref{fig:xgboost} and Table \ref{tab:table3}.
The model's specific input data is as Table \ref{tab:table2}.

\begin{table}[h]
\centering
\caption{Model Performance Evaluation}
\label{tab:table3}
\resizebox{\linewidth}{!}{%
\begin{tblr}{
  cells = {c},
  hline{1,7} = {-}{0.08em},
  hline{2} = {-}{0.05em},
}
Model Names                 & MAPE      & MSE       & MAPE   & R2       \\
Linear Regression           & 0.0185    & 0.1010    & 0.4820 & 0.5966   \\
Support Vector Machine      & 0.0048    & 0.0575    & 0.1727 & 0.7062   \\
Multilayer Perceptron      & 0.1757    & 0.3407    & 6.5909 & -2.3469  \\
Random Forest               & 0.0019    & 0.0256    & 0.0713 & 0.9199   \\
XGBRegressor                & 0.0021    & 0.0282    & 0.0775 & 0.9038   
\end{tblr}
}
\end{table}

\begin{table}[h]
  \centering
  \caption{Model Input}
  \label{tab:table2}
  \begin{tabular}{|c|p{5.5cm}|}
    \hline
    Data Type & Description \\
    
    \hline
    QPS & queries per second (QPS) of the pod \\
    
    \hline
    Performance Metrics & cpu utilization, \\
      & memory usage \\
      & mem\_cache \\
      & mem\_pgfault \\
      & mem\_pgmajfault \\
      & working\_set \\
      & memory\_rss \\
      & net\_recv\_avg \\
      & net\_recv\_packets\_avg \\
      & net\_send\_avg \\
      & net\_send\_packets\_avg \\
      & fs\_read\_avg \\
      & fs\_write\_avg \\

    \hline
    Hardware Events & branch-instructions, ref-cycles, \\
     & branch-misses, bus-cycles, \\
     & cache-misses, cache-references, \\
     & cpu-cycles, instructions, \\
     & alignment-faults, bpf-output, \\
     & cpu-migration, emulation-faults, \\
     & major-faults, minor-faults, \\
     & page-faults, dummy, \\
     & L1-dcache-load-misses, L1-dcache-loads, \\
     & L1-dcache-stores, LLC-load-misses, \\
     & LLC-loads, LLC-store-misses, \\
     & LLC-stores, branch-load-misses, \\
     & branch-loads, dTLB-load-misses, \\
     & dTLB-loads, dTLB-store-misses, \\
     & dTLB-stores, iTLB-load-misses, \\
     & iTLB-loads, node-load-misses \\ 

    \hline
    Scheduling Latency & runqlat[200] \\
     & $runqlat[k] (0 \leq k \leq 199)$ represents the number of occurrences of scheduling latency within the time interval $[k*5, k*5 + 5)$ (unit: ns). Specifically, $runqlat[199]$ indicates the count of occurrences where scheduling latency exceeds or equals 995 ns.\\

    \hline
  \end{tabular}
\end{table}

\subsection{Interference Quantification Module}
\label{subsec:intf-quantification}
Our quantification of interference consists of two parts. The first part assesses the degree of interference on the target host, which we calculate as Equation (\ref{equ:1}).

\begin{equation}
\label{equ:1}
intf_h = w_a \sum_{i=1}^n avg(runqlat^i) + w_b \sum_{j=1}^m avg(runqlat^j)
\end{equation}

In this context, $intf_h$ represents the interference value of node $h$, $n$ denotes the number of online services within node $h$, $runqlat^i$ signifies the scheduling latency array of online service $i$, $m$ signifies the number of offline services within the node, $runqlat^j$ denotes the scheduling latency array of offline service $j$, and $w_a$ and $w_b$ are two weighting factors greater than 1. The $avg()$ function is employed to compute the average scheduling latency, and its calculation method is outlined in (\ref{equ:2}).

\begin{equation}
\label{equ:2}
avg(runqlat)=(\sum_{k=0}^{199} runqlat_k * k * 5) / (\sum_{k=0}^{199} runqlat_k)
\end{equation}

In this context, $runqlat$ represents the data collected for scheduling latency, which is an array with a size of 200. $runqlat_k$ represents the number of occurrences of scheduling latency within the time interval $[k * 5, k * 5 + 5)$ (unit: ns). Specifically, $runqlat_{199}$ indicates the count of occurrences where scheduling latency exceeds or equals 995 ns.

For a submitted pod, we employ a model to predict its potential average scheduling latency when deployed on a node. The description of this model is provided in \cref{subsec:runqlat-module}. That is, the quantification of interference for this pod is as Equation (\ref{equ:3}):

\begin{equation}
\label{equ:3}
intf_p = w_c * model(qps^{pod}, data^n)
\end{equation}

In this context, $intf_p$ signifies the interference value of the pod, $w_c$ represents a weight greater than zero, $model()$ denotes the prediction model, with its output being the predicted average scheduling latency for the pod. $qps^{pod}$ corresponds to the user-specified potential QPS (Queries Per Second) for this pod. Additionally, $data^n$ refers to performance data pertaining to node $n$, with detailed information available in Table \ref{tab:table2}.

\subsection{Scheduling Module}

For an online workload about to be deployed, in order to minimize the interference it may experience post-deployment, we have formulated the following scoring strategy, building upon \cref{subsec:intf-quantification}, as illustrated in Equation (\ref{equ:4}).

\begin{equation}
\label{equ:4}
score_h = (1 - utiliz_h^{cpu}) * (1 - utiliz_h^{mem}) - intf_h - intf_p
\end{equation}

The $score_h$ represents the score of candidate node $h$, and the meanings of $utiliz_h^{cpu}$ and $utiliz_h^{mem}$ are as Equation (\ref{equ:5}) and Equation (\ref{equ:6}).

\begin{equation}
\label{equ:5}
utiliz_h^{cpu} = \frac{cpu_h^{cur} + w_d * cpu^{pod}} {cpu_h^{sum}}
\end{equation}

\begin{equation}
\label{equ:6}
utiliz_h^{mem} = \frac{mem_h^{cur} + w_e * mem^{pod}} {mem_h^{sum}}
\end{equation}

The $cpu_h^{cur}$ and $mem_h^{cur}$ represent the current CPU and memory utilization of the host, while $cpu_h^{sum}$ and $mem_h^{sum}$ denote the total CPU and memory resources of the host. Additionally, $cpu^{pod}$ and $mem^{pod}$ represent the predicted CPU and memory utilization of the pod based on the resource prediction model. $w_d$ and $w_e$ are weight values greater than 1.0.


After receiving the pod submitted by the user and the results from the resource prediction model, the \textit{Scheduling Module} will execute Algorithm 1. The process of Algorithm \ref{alg:1} is as follows. Firstly, it initializes the current best score ($score_{best}$) as negative infinity and the best node number ($node_{selected}$) as -1 (lines 1-2). It also obtains performance data for each node ($nodes_{data}$) through the \textit{Data Collection Module} (line 3). Next, it iterates through each node, extracting $cpu_h^{cur}$, $cpu_h^{sum}$, $mem_h^{cur}$ and $mem_h^{sum}$ from the performance data (lines 4-6). Subsequently, it computes $utiliz_h^{cpu}$ and $utiliz_h^{mem}$ for each node according to Equation (\ref{equ:5}) and Equation (\ref{equ:6}) (line 7). While the goal is to enhance the resource utilization level of nodes, there exists an upper limit. Therefore, it assesses whether $utiliz_h^{cpu}$ and $utiliz_h^{mem}$ exceed their respective thresholds. If they exceed the thresholds, the node is not considered a candidate, and the current iteration proceeds to the next node (lines 8-9). Otherwise, it invokes the \textit{Interference Quantification Module} to obtain $intf_h$ and $intf_p$ for the node and calculates the current node's score using Equation (\ref{equ:4}) (lines 10-11). If the score of the current node surpasses the current best score, the \textit{Scheduling Module} updates the current best score and best node number (lines 12-14). Finally, it returns the computed best node number (line 15).

\begin{algorithm}
\label{alg:1}
\caption{Selection of the best node}
\LinesNumbered

\KwData{$pod$ (user's submission), $cpu^{pod}$ (Resource Prediction Module's output), $mem^{pod}$ (Resource Prediction Module's output)}
\KwResult{node (selected node)}

\BlankLine

$score_{best}$ = $-\infty$\; 
$node_{selected}$ = -1\;
Invoke the \textit{Data Collection Module} and store the results in $nodes\_data$ (an array with data for n nodes)\;

\For{each element i from 1 to n}{
    Fetching $cpu_h^{cur}$, $cpu_h^{sum}$ from $nodes\_data[i]$\;
    Fetching $mem_h^{cur}$, $mem_h^{sum}$ from $nodes\_data[i]$\;
    Calculate $node_i$'s $utiliz_h^{cpu}$ and $utiliz_h^{mem}$\;
    \If{($utiliz_h^{cpu} > 0.70$) \  or \  ($utiliz_h^{mem} > 0.80$)}{
        Continue with the next iteration\;
    }

    Invoke the \textit{Interference Quantification Module} to retrieve $intf_h$ and $intf_p$ of $node_i$\;
    Calculate $node_i$'s $score_h$\;

    \If{$score_h > score_{best}$}{
        $score_{best} = score_h$\;
        $node_{selected} = i$\;
    }
}

return $node_{selected}$\;

\end{algorithm}

\section{Experimental Results} \label{Experimental Results}
To assess the effectiveness of our approach, we conducted experiments and documented the experimental results. In this section, we will provide an overview of the experimental procedure and present the obtained results.

\subsection{Benchmark Selection}
In cloud environments, there are several typical application types for both online and offline workloads. Typical online workload types include Web Serving, Web Search, Media Streaming\footnote{https://github.com/parsa-epfl/cloudsuite/blob/main/docs/benchmarks/media-streaming.md}, and Data Caching\footnote{https://github.com/parsa-epfl/cloudsuite/blob/main/docs/benchmarks/data-caching.md}. Typical offline workload types include Graph Analysis\footnote{https://github.com/parsa-epfl/cloudsuite/blob/main/docs/benchmarks/graph-analytics.md} and In-Memory Analytics.

\subsection{Metric Collection}
We utilize Prometheus to collect runtime resource utilization data for each application and node. Prometheus is an open-source monitoring solution used for collecting and aggregating metrics as time-series data.

We employ Perf to collect hardware-level metrics from the host. Perf is a performance profiling tool that is built into the Linux kernel source tree. It operates on the principle of event sampling and is based on performance events, supporting performance profiling for both processor-related and operating system-related performance metrics.

We utilize eBPF to collect scheduling latency data. It's worth noting that eBPF comes with a tool for collecting scheduling latency data, which aggregates data in exponentially growing intervals, such as [0,2), [2, 4), …, [64, 128). However, this non-linear approach lacks intuitiveness and cannot accurately calculate the average scheduling latency, resulting in certain limitations. To address this, we have made improvements by collecting data in intervals of 5 nanoseconds each, totaling 200 intervals. The last interval is used to store the count of scheduling latencies equal to or exceeding 995 nanoseconds.

\subsection{Application Performance Testing}
In order to closely approximate real-world scenarios when sending query requests for performance testing, we referenced the open-source dataset (Cluster-trace-v2018\footnote{https://github.com/alibaba/clusterdata/blob/master/cluster-trace-v2018/trace\_2018.md}) provided by Alibaba. The Cluster-trace-v2018 dataset encompasses the changes in the status data of approximately 4,000 machines over an 8-day period. During the testing phase, we generated requests with dynamically changing QPS over time. The variation in QPS mimicked the resource utilization trends observed in the dataset for applications as time progressed. For example, if we aimed for an average QPS of approximately 300, the actual QPS fluctuated dynamically around this target value.

\subsection{Experimental Result}

We have developed a program for testing the scheduler, which can submit a pod after a random time interval. In addition to the scheduler implemented according to our proposed interference-aware container orchestration (ICO), we also tested Round Robin Scheduling (RR), High Utilization Priority Scheduling (HUP) and Low QPS Priority Scheduling (LQP).

The RR is a straightforward and equitable scheduling approach. It cyclically assigns pending pods to various nodes, ensuring that each node has an opportunity to execute tasks.

The HUP is an effective method for enhancing node resource utilization derived from \cite{Lu2023}. We have made certain modifications to this method. The scoring mechanism for this algorithm is shown as Equation (\ref{equ:7}).

\begin{equation}
\label{equ:7}
HUPscore_h = utiliz_h^{cpu} * utiliz_h^{mem} - intf_h - intf_p
\end{equation}

In this context, the meanings of $utiliz_h^{cpu}$, $utiliz_h^{mem}$, $intf_h$ and $intf_p$ are consistent with those defined in Equation (\ref{equ:1}), (\ref{equ:3}), (\ref{equ:5}) and (\ref{equ:6}).

The LQP refers to the approach in which, when a pod needs to be scheduled, the total sum of online workloads QPS on each node is calculated. The node with the lowest QPS sum is selected as the target node.

After the experiments concluded, we computed the average response time, 90th percentile response time, and 99th percentile response time for all pods under various scheduling algorithms. The results are illustrated in Figure \ref{fig:result1} below.


  
\begin{figure}[htbp]
  \centering 
  \includegraphics[width=\linewidth]{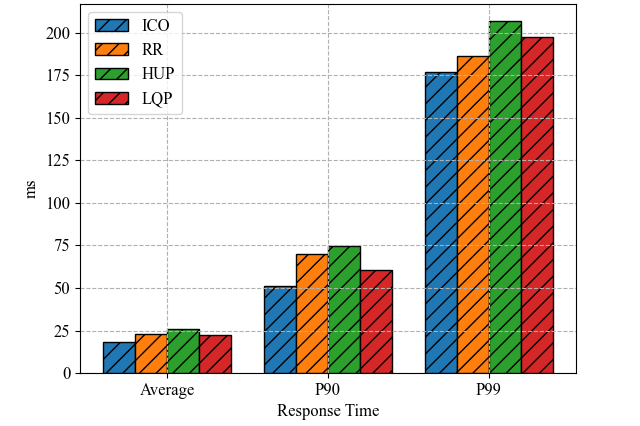}
  \caption{Comparison of results from different schedulers}
  \label{fig:result1}
\end{figure}

\begin{figure}[h]
  \centering
  \setlength{\abovecaptionskip}{-0.2cm} 
  \begin{minipage}{0.47\linewidth} 
    \centering
    \includegraphics[width=\linewidth]{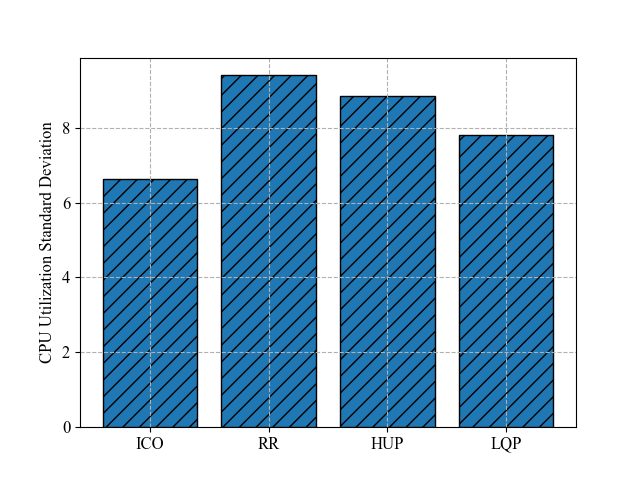}
    \caption{Cluster nodes' CPU utilization standard deviation}
    \label{fig:result-cpustd}
  \end{minipage}%
    \begin{minipage}{0.06\linewidth}
    \centering
    \qquad
  \end{minipage}%
  \begin{minipage}{0.47\linewidth} 
    \centering
    \includegraphics[width=\linewidth]{result_cpustd_latest.png}
    \caption{Cluster nodes' MEM utilization standard deviation}
    \label{fig:result-memstd}
  \end{minipage}
\end{figure}

Additionally, we collected data on CPU utilization and memory utilization for each node during the experimental process. We aimed to achieve a balanced resource utilization across all nodes; therefore, we further calculated the variance in CPU utilization and memory utilization among the nodes. The results are presented in Figure \ref{fig:result-cpustd} and Figure \ref{fig:result-memstd}.

In terms of average response time, 90th percentile response time, and 99th percentile response time for online workloads, our method reduces these metrics by 29.4\%, 31.4\%, and 14.5\%, respectively, compared to HUP. In comparison to the RR, our method reduces them by 20.1\%, 27.2\%, and 5.1\%, respectively. When compared to LQP, our method reduces these metrics by 16.7\%, 15.4\%, and 10.3\%, respectively.

Regarding cluster utilization, the standard deviation of CPU utilization among nodes was 6.63 (in contrast, the results for RR, HUP and LQP were 9.40, 8.84 and 7.81, respectively), while the standard deviation of MEM (memory) utilization was 6.53 (compared to 18.24, 12.76 and 18.93 for RR, HUP and LQP, respectively). This study provides novel insights and methodologies for optimizing the joint deployment of online and offline workloads, facilitating their hybrid deployment.



\section{Conclusions and Future Work} \label{Conclusions and Future Work}


In this paper, we introduce the concept of scheduling latency and propose a method for quantifying interference and scheduling algorithms based on this metric. Through experimental validation, our approach has demonstrated promising results. 

In our future work, we plan to further optimize the scheduling latency metric with a focus on several key aspects.
First, in this study, we confirmed a strong correlation between scheduling latency and application response times. Therefore, we intend to explore adjustments in process priorities and other relevant factors to ensure that the scheduling latency of critical processes remains within acceptable bounds, thus guaranteeing the desired application response times.
Second, our current efforts have primarily concentrated on the initial scheduling phase. In the upcoming phases, we will dedicate our efforts to refining the design of secondary scheduling processes.
Last, in our current approach, resource allocation is fixed after the initial scheduling phase. In the future, we will integrate the scheduling latency metric into the development of a dynamic scaling and resource allocation mechanism to adaptively adjust resource allocation based on changing workload conditions.

\bibliographystyle{IEEEtran}
\bibliography{b.bib}

\end{document}